\RequirePackage[hyphens]{url}
\documentclass[aip,jcp,twocolumn,showpacs,bibnotes,reprint,superscriptaddress,floatfix]{revtex4-1}

\usepackage{hyperref,url}
\usepackage{amsmath}
\usepackage[capitalise]{cleveref}
\hypersetup{breaklinks,colorlinks=true,linkcolor=blue,citecolor=blue,urlcolor=blue,filecolor=blue}
\usepackage{graphicx}
\usepackage{dcolumn}
\newcommand{\bk}{\mathbf{k}}

\newcommand{\bQ}{\mathbf{Q}}

\newcommand{\br}{\mathbf{r}}
\newcommand{\bR}{\mathbf{R}}
\newcommand{\bx}{\mathbf{x}}

\newcommand{\hh}{\hat{H}}

\newcommand{\hc}{\hat{c}}
\newcommand{\hcd}{\hat{c}^{\dagger}}

\begin{document}

\title{Accelerating Auxiliary-Field Quantum Monte Carlo Simulations of Solids with Graphical Processing Units}
\author{Fionn D. Malone}
\affiliation{Quantum Simulations Group, Lawrence Livermore National Laboratory, Livermore, California 94550, USA}
\author{Shuai Zhang}
\affiliation{Quantum Simulations Group, Lawrence Livermore National Laboratory, Livermore, California 94550, USA}
\affiliation{Laboratory for Laser Energetics, University of Rochester, 250 E River Rd, Rochester, NY 14623, USA}
\author{Miguel A. Morales}
\email{moralessilva2@llnl.gov}
\affiliation{Quantum Simulations Group, Lawrence Livermore National Laboratory, Livermore, California 94550, USA}

%%%%%%%%%%%%%%%%%%%%%%%%%%%%%%%%%%%%%%%%%%%%%%%%%%%%%%%%%%%%%%%%%%%%%
%% The document title should be given as usual. Some journals require
%% a running title from the author: this should be supplied as an
%% optional argument to \title.
%%%%%%%%%%%%%%%%%%%%%%%%%%%%%%%%%%%%%%%%%%%%%%%%%%%%%%%%%%%%%%%%%%%%%

%%%%%%%%%%%%%%%%%%%%%%%%%%%%%%%%%%%%%%%%%%%%%%%%%%%%%%%%%%%%%%%%%%%%%
%% Some journals require a list of abbreviations or keywords to be
%% supplied. These should be set up here, and will be printed after
%% the title and author information, if needed.
%%%%%%%%%%%%%%%%%%%%%%%%%%%%%%%%%%%%%%%%%%%%%%%%%%%%%%%%%%%%%%%%%%%%%
%\abbreviations{IR,NMR,UV}
%\keywords{American Chemical Society, \LaTeX}

%%%%%%%%%%%%%%%%%%%%%%%%%%%%%%%%%%%%%%%%%%%%%%%%%%%%%%%%%%%%%%%%%%%%%
%% The manuscript does not need to include \maketitle, which is
%% executed automatically.
%%%%%%%%%%%%%%%%%%%%%%%%%%%%%%%%%%%%%%%%%%%%%%%%%%%%%%%%%%%%%%%%%%%%%
\begin{abstract}
We outline how auxiliary-field quantum Monte Carlo (AFQMC) can leverage graphical processing units (GPUs) to accelerate the simulation of solid state sytems.
By exploiting conservation of crystal momentum in the one- and two-electron integrals we show how to efficiently formulate the algorithm to best utilize current GPU architectures. 
We provide a detailed description of different optimization strategies and profile our implementation relative to standard approaches, demonstrating a factor of 40 speed up over a CPU implementation.
With this increase in computational power we demonstrate the ability of AFQMC to systematically converge solid state calculations with respect to basis set and system size by computing the cohesive energy of Carbon in the diamond structure to within 0.02 eV of the experimental result.
\end{abstract}
\maketitle

%\begin{tocentry}

%Insert graphical entry here.

%\end{tocentry}


\section{Introduction}
Auxiliary-field quantum Monte Carlo\citep{zhang_cpmc,Zhang_phaseless} (AFQMC) has emerged of late as one of the most accurate approaches to the electronic structure problem\citep{motta_review,motta_hydrogen_benchmark_2017,williams_transition_metal_simons,motta_hydrogen_2019}.
Capable of treating challenging real\citep{alsaidi_trans_06,purwanto_chromium_dimer,purwanto_excited_c2,ma_downfold_2015,purwanto_downfolding_jctc,shee_correlated_sampling,shee_transition_metals,Shee_ST_2019,zhang_nio,lee_st_afqmc_2020} and model systems\citep{2d_hubbard_benchmark,lee_2019_UEG,qin_hubbard_sc_2019}, its accuracy for single reference problems lies somewhere between coupled cluster with  singles and doubles (CCSD) and CCSD with perturbative triples (CCSD(T))\citep{bartlett2007coupled}.
AFQMC has the added advantage of a favorable scaling with system size, between $\mathcal{O}(N^3)$-$\mathcal{O}(N^4)$, %when using Gaussian basis sets
\citep{motta_review,purwanto_cholesky,pw_footnote} in contrast to $\mathcal{O}(N^6)$ or $\mathcal{O}(N^7)$ for canonical CCSD and CCSD(T) respectively. 
In addition, unlike most traditional wavefunction-based quantum chemical methods, AFQMC uses Monte Carlo methods to stochastically solve the many-electron Schr{\"o}dinger equation.
This stochastic sampling naturally allows for the exploitation of massively parallel supercomputing resources.

Despite these apparent advantages, the widespread adoption of AFQMC has been hindered in part because of a large computational prefactor that masks the method's favorable scaling with system size.
Although algorithmic advances, such as the use of tensor hyper-contraction\citep{malone_isdf,motta_thc} and by explicitly exploiting symmetry in the two-electron integrals\citep{motta_kpoint} have shown that it is possible to reduce the computational cost and memory overhead by an order of magnitude, the time to solution of the method can often be prohibitive for researchers without access to large supercomputing resources.
This is particularly problematic for solid state applications, where systems with thousands of orbitals are necessary to obtain physically meaningful results\citep{Zhang_phaseless,malone_isdf}.

Fortunately, recent years have seen a rapid growth in the development and use of graphical processing units (GPUs) to accelerate computationally intensive tasks\citep{Lecun_deep_learning_2015}.
As a result, a mature and relatively user friendly software stack is becoming available for most of the common operations needed for electronic structure codes.
This has significantly lowered the effort required to write efficient GPU code.
In particular, pioneered by Nvidia and now followed by library developers and several GPU vendors, batched extensions to BLAS and LAPACK libraries are available which enable the concurrent execution of a large number of small matrix operations, leading to good performance in GPUs for algorithms which would otherwise struggle to extract the computational power of these new, high-throughput architectures.
This paradigm shift in computing is now driving the development of the next generation of supercomputers and the electronic structure community needs to adapt in order to use them\citep{Kothe_exascale_2020}.
GPUs are already being leveraged in Hartree--Fock and density functional theory calculations \citep{Ufimtsev_gpuqhcem_2008,Ufimtsev_gpu1_2008,Ufimtsev_gpu2_2009,Ufimtsev_gpu3_09,jia_pwgpu_2013}as well as in correlated methods\citep{Anderson_qmc_gpu_2007,Vogt_rimp2_gpu_2008,Meredith_qmcgpu_2009,Olivares-Amaya_qchem_gpu_2010,Watson_qchem_gpu_2010,Ma_ccsdt_gpu_2010,Ma2_ccsdt_gpu_2011,Esler_gpu_2012,qmcpack} and molecular dynamics simulations\citep{stone_gpu_mol_model_2010}.
Unfortunately progress has been slow due to the need for custom-made code for the GPUs in most situations, whose different architecture
and computing capability (large number of concurrent and independent SIMD engines) makes codes designed for multi-core CPUs usually very slow.
Fortunately, as the AFQMC algorithm is largely reliant on dense linear algebra operations, it is ideally suited for leveraging accelerator devices.
Indeed, it has already been shown in Ref.\citenum{Shee_gpu_2018} that significant speedups of AFQMC calculations of molecules on GPUs are possible.
In this paper we outline how similar gains can be made in solid state systems, particularly if crystal momentum conservation is exploited\citep{motta_kpoint}.

This paper is outlined as follows.
We first review the AFQMC method and different implementation strategies for ab-initio systems.
Next in we outline how to efficiently implement AFQMC to best use modern GPUs. 
Finally, we outline our results for the cohesive energy of carbon in the diamond structure and finish with some closing remarks about future prospects for the method. 

%Developments in computing technologies enable faster supercomputers with GPU interfaces and new programming tools (e.g., CUDA). AFQMC is particularly well suited for GPU with careful implementations. GPU now allow us to study systems with thousands of basis functions (in this paper up to ~7200) in reasonable time, dramatically expanding the reach of AFQMC in complex systems.

%implement an efficient algorithm and develop AFQMC codes that works on GPU computers. 
\section{Methods \label{sec:methods}}
In this section we briefly review the basics of ph-AFQMC and its application to ab-initio systems.
In what follows we work in a basis of $M$ orthogonalized ortitals denoted by $p,q,r,s$ for arbitrary basis functions with $a,b$ used to denote occupied orbitals. For supercells we consider systems with $N$ electrons. When working in the $k$-point representation lower case letters will be used for the average number of bands per $k$-point ($m$) and the number of electrons per $k$-point ($n$). Bold face symbols will be used to denote vectors and matrices ($\mathbf{V}$) with their elements given by, for example, $V_{pq}$.
\subsection{Introduction to AFQMC\label{sec:intro_afqmc}}
In this work we are interested in determining the ground state properties of the many-electron Hamiltonian:
\begin{align}
    \hat{H} &= \sum_{pq\sigma} h_{pq} \hcd_{p\sigma}\hc_{q\sigma} + \frac{1}{2}\sum_{pqrs\sigma\sigma'}v_{pqrs}  \hcd_{p\sigma}\hcd_{q\sigma'}\hc_{s\sigma'}\hc_{r\sigma}+E_{II},\label{eq:hamil}\\
            &= \hh_1 + \hh_2 + E_{II},
\end{align}
where $E_{II}$ is the ion-ion energy contribution, and $\hcd_{p\sigma}$ and $\hc_{p\sigma}$ create and annihilate an electron in some single-particle state of spin $\sigma$, $|p\sigma\rangle$. The matrix elements of the one- and two-body parts of the Hamiltonian are given (in Hartree atomic units) as
\begin{equation}
h_{pq} = \int d\br \  \varphi_{p}^*(\br)\left(-\frac{1}{2}\hat{\nabla}_\br^2 - \sum_I \frac{Z}{|\br-\bR_I|}\right)\varphi_{q}(\br),
\end{equation}
where $\langle \br | p\rangle = \varphi_{p}(\br)$ is some single-particle orbital, $Z$ is the ionic charge and $\bR_I$ is the location of ion $I$.
The electron-repulsion integrals (ERIs) are in turn given by:
\begin{equation}
    v_{pqrs} = \int \int d\br \ d\br' \ \varphi^*_p(\br)\varphi^*_q(\br')\frac{1}{|\br-\br'|}\varphi_{r}(\br)\varphi_s(\br')\label{eq:four_ix}.
\end{equation}

One way to find the ground state wavefunction, $|\Psi_0\rangle$, of the Hamiltonian given in \cref{eq:hamil} is through imaginary time projection:
\begin{align}
|\Psi_0\rangle \propto &\lim_{n\rightarrow\infty} \left(e^{-\Delta\tau\hat{H}}\right)^{n}|\Psi_I\rangle, \\ \label{eq:projection}
                   =   & \lim_{n\rightarrow\infty} |\Psi^{(n)}\rangle,
\end{align}
where $\Delta\tau$ is the timestep and $|\Psi_I\rangle$ is some initial state satisfying $\langle \Psi_I | \Psi_0\rangle \ne 0$.
In AFQMC, as is common to many projector QMC methods, we first employ a (symmetrized) Suzuiki--Trotter approximation to write the imaginary time evolution operator as
\begin{equation}
    e^{-\Delta\tau\hat{H}} \approx e^{-\frac{\Delta\tau}{2} \hat{H}_1} e^{-\Delta\tau \hat{H}_2} e^{-\frac{\Delta\tau}{2} \hat{H}_1}.
\end{equation}
We next write the two-body Hamiltonian as a sum of squares of one-body operators
\begin{align}
    \hat{H}_2 &=  \hat{v}_0 - \frac{1}{2} \sum_{\gamma} \hat{v}_{\gamma}^2,\label{eq:hsq}\\
              &= \hat{v}_0 + \hat{H}_2'
\end{align}
where
\begin{equation}
\hat{v}_0 = -\frac{1}{2}\sum_{pq\sigma} \left(\sum_r v_{prrq}\right)\hcd_{p\sigma}\hc_{q\sigma}.
\end{equation}
The two-body propagator can now be written in terms of one-body propagators only using the Hubbard--Stratonovich transformation\citep{hubbard_strat}
\begin{align}
e^{\frac{\Delta\tau}{2}\hat{v}_\gamma^2} & = \int \frac{d x_\gamma}{\sqrt{2\pi}} e^{-\frac{x_\gamma^2}{2}} 
e^{\sqrt{\Delta\tau} x_\gamma \hat{v}_\gamma},
\label{eq:hs}
\end{align}
so that the projection to the ground state can be achieved iteratively  via
\begin{equation}
    |\Psi^{(n+1)}\rangle = \int d \bx p(\bx) \hat{B}(\mathbf{x}) |\Psi^{(n)}\rangle\label{eq:iterate},
\end{equation}
where
\begin{equation}
\hat{B}(\mathbf{x}) = e^{-\frac{\Delta\tau}{2} \hat{H}'_1} e^{\sqrt{\Delta\tau} \mathbf{x}\cdot\hat{\mathbf{v}}} e^{-\frac{\Delta\tau}{2} \hat{H}'_1},\label{eq:propagator}
\end{equation}
and $\hat{H}_1'=\hat{H}+\hat{v}_0$.
In practice, the multi dimensional integral in \cref{eq:iterate} is evaluated using Monte Carlo methods.
That is, we sample a statistical representation of the wavefunction using a finite ensemble of random walkers 
\begin{equation}
    |\Psi^{(n)}\rangle = \sum_\alpha^{N_w} w_\alpha^{(n)} |\phi_\alpha^{(n)} \rangle,
\end{equation}
where $N_w$ is the total number of walkers.
At each time step we draw a normally distribution auxiliary field, $\bx$, construct $\hat{B}(\bx)$ and apply this to the walker's Slater determinant $|\phi\rangle$, yielding an updated single Slater determinant\citep{thouless_theorem,thouless_theorem_2}.
This `free-projection' AFQMC is limited by a phase problem which arises due to the generally complex propagator $\hat{B}(\mathbf{x})$.
In this work we instead use the phaseless AFQMC method\citep{Zhang_phaseless} to overcome this issue at the expense of introducing a systematic bias in our results.

Practically, phaseless AFQMC (ph-AFQMC) amounts to first performing an importance sampling transformation so that walkers undergo the modified propagation
\begin{equation}
    w_\alpha^{(n+1)}|\phi_\alpha^{(n+1)}\rangle =   \left[I(\bx,\bar{\bx},|\phi^{(n)}\rangle) \hat{B}(\bx-\bar{\bx})\right]w_\alpha^{(n)}|\phi^{(n)}\rangle, \label{eq:iterate_imp}
\end{equation}
where
\begin{equation}
I(\bx,\bar{\bx},|\phi\rangle) = \frac{\langle \psi_T|\hat{B}(\bx- \bar{\bx}) |\phi\rangle}{\langle \psi_T|\phi\rangle} e^{\bx\cdot\bar{\bx}-\frac{\bar{\bx}\cdot\bar{\bx}}{2}},
\end{equation}
is the importance function, $\bar{\bx}$ is the `force-bias' shift
given by
\begin{equation}
    \bar{x}_\gamma = -\sqrt{\Delta\tau} \frac{\langle \psi_T | \hat{v}_\gamma | \phi\rangle}{\langle \psi_T | \phi \rangle },
\end{equation}
and $|\psi_T\rangle$ is a trial wavefunction.
To control the phase problem the walker's weights is updated in `hybrid' form
\begin{equation}
    w_\alpha^{(n+1)} = |I(\bx,\bar{\bx},|\phi_\alpha^{(n)}\rangle)|\times \max \left(0, \cos \Delta \theta\right) w_\alpha^{(n)},
\end{equation}
where the phase is defined as
\begin{equation}
    \Delta \theta = \arg\left(\frac{\langle \psi_T|\hat{B}(\bx- \bar{\bx}) |\phi\rangle}{\langle \psi_T|\phi\rangle}\right).
\end{equation}
This procedure kills walkers whose phase changes by $\pi/2$ in any one step and prevents the accumulation of weight near the origin in the complex plane which would otherwise render the method impractical\citep{zhang_julich_2013}.
The trial wavefunction enforces the `phaseless' constraint, which produces exact results if $|\Psi_T\rangle=|\Psi_0\rangle$.
Although approximate, ph-AFQMC has been applied successfully to compute ground state\citep{motta_back_prop,motta_forces}, and excited state\citep{ma_excited_state_2013,motta_itcf_1,motta_itcf_2} properties of a variety of molecules and solids, showing often remarkable accuracy with very simple single determinant trial wavefunctions\citep{lee_st_afqmc_2020}.
In what follows we will refer to ph-AFQMC as AFQMC for brevity.

\subsection{Standard Representation \label{sec:standard}}

Central to the practical application of the AFQMC algorithm is the factorization of the ERIs.
In the standard approach we use a modified Cholesky decomposition\citep{modified_chol_1,modified_chol_2,modified_chol_3} to write
\begin{equation}
    v_{pqrs} \approx \sum_\gamma^{N_\gamma} L_{pr,\gamma} L^*_{sq,\gamma},
\end{equation}
where $N_\gamma = n_\gamma M$ is the number of Cholesky vectors necessary to reproduce the ERIs to within a given threshold, and $M$ is the number of single-particle basis functions.
For typical systems\citep{motta_review} $n_\gamma$ is in the range of $5-10$.
With this factorization we next introduce the Hubbard--Stratonovich `potentials',
\begin{align}
    \hat{v}_{\gamma\pm} &= c_\pm \sum_{pr\sigma} \left(\frac{L_{pr,\gamma}\pm L^*_{rp,\gamma}}{2}\right)\hcd_{p\sigma}\hc_{r\sigma} \\
    &=  c_\pm \sum_{pr\sigma}\left[L_{\pm}\right]_{pr,\gamma}\hcd_{p\sigma}\hc_{r\sigma}, \\
\end{align}
where $c_+ = 1$ and $c_-=i$, so that
\begin{equation}\label{eq:h2_mf}
         \hh_2' = -\frac{1}{2} \sum_{\gamma\pm} \hat{v}_{\gamma\pm}^2.
    %\begin{split}
         %\hh_2' = -\frac{1}{2} \sum_{\gamma\pm} \left[ &\left(\hat{v}_{\gamma\pm}-\bar{v}_{\gamma\pm}\right)^2  \right.\\
         %                                       & \left. +2\hat{v}_{\gamma\pm}\bar{v}_{\gamma\pm} -\bar{v}^2_{\gamma\pm} \right] 
    %\end{split}
%    \begin{split}
%        & \left(\hat{v}_{\gamma\pm}-\bar{v}_{\gamma\pm}\right)^2   \\
        %& \left. + 2\hat{v}_{\gamma\pm}\bar{v}_{\gamma\pm}\right]
           %- \bar{v}_{\gamma\pm}^{2} \right]
%    \end{split}
\end{equation}
%where we introduced a mean-field shift to the potentials 
%\begin{equation}
%    \bar{v}_{\gamma\pm} = \frac{\langle \psi_T | \hat{v}_{\gamma\pm} |\psi_T \rangle}{\langle \psi_T|\psi_T\rangle},
%\end{equation}
%to help reduce the severity of the phase problem in AFQMC\citep{purwanto_back_prop}.
%With these definitions we make the replacement $\hat{v}_{\gamma\pm}'=\hat{v}_{\gamma\pm}-\bar{v}_{\gamma\pm}$ and absorb the one-body and constant term in \cref{eq:h2_mf} into the definition of $\hat{H}_1'$ in \cref{eq:propagator}.
Finally, the force bias shift is given by
\begin{equation}
\bar{x}_{\gamma\pm} = -\sqrt{\Delta\tau}c_\pm\sum_{pr\sigma}[L_{\pm}]_{pr,\gamma}G_{pr}^{\sigma},
\label{eq:fb}
\end{equation}
where we have identified the Green's function
\begin{align}
    G^{\sigma}_{pr} &= \frac{\langle\psi_T|\hcd_{p\sigma}\hc_{r\sigma}|\phi\rangle}{\langle\psi_T|\phi\rangle}\\
                           &= \left[U_{\sigma}(A_\sigma ^{\dagger}U_{\sigma})^{-1}A_{\sigma}^{\dagger}\right]_{rp}\\
    &=\left[A_\sigma^*(U_{\sigma}^T A_\sigma^*)^{-1}U^{T}_{\sigma}\right]_{pr}\label{eq:gf},
\end{align}
where $U_\sigma$ and $A_\sigma$ are the $M\times N_\sigma$  matrices of orbital coefficients for the walker $|\phi\rangle$ and trial wavefunction $|\psi_T\rangle$ respectively.
It is advantageous at this point to introduce `half-rotated' Green's functions respectively and Hubbard--Stratonivich potentials, 
\begin{align}
    G_{p r}^\sigma &= \sum_a^{N_\sigma} [A^{*}_\sigma]_{pa} \mathcal{G}_{ar}^{\sigma},\\
    \mathcal{G}_{ar}^{\sigma} &= [(U_{\sigma}^T A_\sigma^*)^{-1} U^{T}_{\sigma}]_{ar},\label{eq:gfhrot}
\end{align}
\begin{equation}
    \left[\mathcal{L}_{\pm}\right]_{ar,\gamma}^{\sigma} = \sum_s \left[A^{*}_\sigma\right]_{ar}\left[L_{\pm}\right]_{ar,\gamma},
\end{equation}
so that\cite{motta_review}
\begin{equation}
    \bar{x}_{\gamma\pm} = -\sqrt{\Delta\tau}c_\pm \sum_{ak\sigma}\left[\mathcal{L}_{\pm}\right]^{\gamma}_{ak\sigma}\mathcal{G}_{a\sigma k\sigma},
\end{equation}
bringing the cost of computing the force-bias down from $\mathcal{O}(N_\gamma M^2)$ to $\mathcal{O}(N_\gamma NM)$ since $\mathcal{L}_{\pm}$ can be computed once at the start of the simulation at the cost of $\mathcal{O}(N_\gamma NM^2)$ operations\citep{motta_review}.

With the force bias and Hubbard--Stratonovich potentials we can construct the matrix
\begin{equation}
    V^{\mathrm{HS}}_{pr} = \sqrt{\Delta\tau} \sum_{\gamma_\pm} c_\pm  [L_\pm]_{pr,\gamma} (x_{\gamma\pm}-\bar{x}_{\gamma\pm})\label{eq:vhs},
\end{equation}
to form the interaction part of the propagator.
The matrix exponential is evaluation as a truncated (typically fourth order) Taylor series expansion\citep{Zhang_phaseless}. 
The cost of propagating a walker is thus $\mathcal{O}(n_\gamma M^3)$  for forming $\mathbf{V}^{\mathrm{HS}}$ and $\mathcal{O}(M^2 N)$ for applying the exponential. 

Finally, the mixed estimate for the local energy at a given timestep $n$ is given by
\begin{align}
    E^{(n)}_\mathrm{mixed} &= \frac{\langle \psi_T | \hat{H} |\Psi^{(n)}\rangle}{\langle \Psi_T| \Psi^{(n)}\rangle} \\
                       &= \frac{\sum_\alpha w^{(n)}_\alpha E_L[\phi^{(n)}_\alpha]}{\sum_{\alpha}w^{(n)}_\alpha},
\end{align}
where
\begin{equation}
    \begin{split}
        E_L[\phi] &= \sum_{pq\sigma} h_{pq}G^{\sigma}_{pq} + \\
        &\frac{1}{2}\sum_{pqrs\sigma\sigma'} v_{pqrs} \left(G^{\sigma}_{p
                r}G^{\sigma'}_{qs}-\delta_{\sigma\sigma'}G^{\sigma}_{p
                s}G^{\sigma'}_{qr}\right)\label{eq:local_energy} \\ 
        &= E_{1B} + E_C + E_{X},
    \end{split}
\end{equation}
is the walker's local energy.
To evaluate \cref{eq:local_energy} efficiently we first define 
\begin{align}
    X_\gamma^{\sigma} &= \sum_{ar} \mathcal{L}^{\sigma}_{ar,\gamma} \mathcal{G}_{ar}^{\sigma},\\
    \bar{X}_\gamma^{\sigma} &= \sum_{bs} \bar{\mathcal{L}}^{\sigma}_{bs,\gamma} \mathcal{G}_{bs}^{\sigma},
\end{align}
where we have similarly defined the half-rotated Cholesky vectors 
\begin{align}
\mathcal{L}^{\sigma}_{ar,\gamma} &= \sum_{p} L_{pr,\gamma} [A_{\sigma}^*]_{pa},\\
\bar{\mathcal{L}}^{\sigma}_{bs,\gamma} &= \sum_{q} L_{sq,\gamma}^* [A^{*}_\sigma]_{qb},
\end{align}
so that the Coulomb energy can be evaluated as
\begin{equation}
E_C = \frac{1}{2} \sum_{\gamma\sigma,\sigma'} X_\gamma^{\sigma} \bar{X}_\gamma^{\sigma'},
\end{equation}
at the cost of $\mathcal{O}(n_\gamma M)$ since the half-rotated Cholesky vectors can be constructed once at the beginning on the simulation at the cost of $\mathcal{O}(n_\gamma NM^2)$.
For the exchange energy we form
\begin{align}
    T_{ab,\gamma}^{\sigma} = \sum_r \mathcal{L}^\sigma_{ar}\mathcal{G}^\sigma_{br}\\
    \bar{T}_{ba,\gamma}^{\sigma} = \sum_s \bar{\mathcal{L}}^{\sigma'}_{bs}\mathcal{G}^{\sigma'}_{as}
\end{align}
so that
\begin{equation}
    E_{X} = -\frac{1}{2} \sum_{ab\gamma\sigma} T_{ab,\gamma}^{\sigma} \bar{T}_{ba,\gamma}^{\sigma},
\end{equation}
at the cost of $\mathcal{O}(n_\gamma N^2 M^2) + \mathcal{O}(n_\gamma N^2 M)$.
Note that one can precompute a half rotated integral tensor which reduces the complexity of the energy evaluation by a factor of at least $n_\gamma$, at the cost of a $\mathcal{O}(N^2M^2)$ memory overhead\citep{zhang_nio,Shee_gpu_2018}. This approach is used in QMCPACK for periodic systems with sparsity. For systems with low degrees symmetry or sparsity, we have found it better to use the direct approach outlined above in order to avoid the prohibitive memory cost which is a significant limitation on GPUs. 

As described, the standard approach has a cubic memory footprint and a quartic computational overhead for the energy evaluation which will dominate the calculation for large system sizes.
However,often many elements of the Cholesky integrals are often either identically zero by symmetry or can be efficiently screened to increase the sparsity of the tensors.
In particular, for periodic systems, conservation of crystal momentum increases the sparsity of the integrals by a factor of $N_k$, where $N_k$ is the number of $k$-points used to sample the Brillouin zone\citep{motta_kpoint}.
In prior work we accounted for this by using sparse linear algegra, which naturally exposes the sparsity of the integrals\citep{zhang_nio}.
As we will see in later sections, dense linear algebra is best suited to modern GPU architectures.
By explicitly incorporating $k$-point symmetry in the two-electron integrals we can reformulate the AFQMC algorithm to involve many small dense operations, thus increasing the efficiency of the method tremendously.

\subsection{$k$-Point Representation \label{sec:kpoint}}

For periodic systems with lattice translational symmetry, significant reductions in memory usage and computational costs can be achieved by representing the Hamiltonian explicitly in terms of band and $k$-point indices\citep{motta_kpoint}. Notice that while it is also possible to take advantages of point group symmetries in periodic systems, which would lead to further reductions given roughly by the order of the symmetry group, in this article we limit the discussion to lattice translational symmetry. In the explicit $k$-point representation, the one- and two-body parts of the Hamiltonian take the form: 
\begin{align}
    \hh_1 & =  \sum_{\textbf{k}pq\sigma} h_{(\textbf{k}p),(\textbf{k}q)} \hcd_{(\textbf{k}p)\sigma}\hc_{(\textbf{k}q)\sigma} ,\label{eq:kp_h1} \\
   \hh_2 & =  \frac{1}{2} \sum_{\substack{\gamma\textbf{Q}\textbf{k}\textbf{k}' \\ pqrs\sigma\sigma'}} L^{\textbf{Q},\textbf{k}}_{pr,\gamma} {L^{\textbf{Q},\textbf{k}'}_{sq,\gamma}}^{*} \nonumber \\ 
  & \ \ \ \ \  \hcd_{(\textbf{k}p)\sigma}\hcd_{(\textbf{k}'-\textbf{Q}q)\sigma'}\hc_{(\textbf{k}'s)\sigma'}\hc_{(\textbf{k}-\textbf{Q}r)\sigma},\label{eq:kp_h2} 
\end{align}
where $\textbf{k}$, $\textbf{k}'$ and $\textbf{Q}$ are vectors in the first Brillouin zone\cite{convention}. 
The one-body Hamiltonian is block diagonal in $\textbf{k}$ and in \cref{eq:kp_h2} we have used the fact that momentum conservation requires that $(\textbf{k}_p - \textbf{k}_r + \textbf{k}_q - \textbf{k}_s) = \textbf{G}$, $\textbf{G}$ being some vector in the reciprocal lattice of the simulation cell. The convention in the notation of the Cholesky matrix $L^{\textbf{Q},\textbf{k}}_{pr,\gamma}$ is defined by $\textbf{k}_r = \textbf{k}_p - \textbf{Q}$, so the vector $\textbf{k}$ labels the \textit{k}-point of the first band index, $\textit{p}$, while the \textit{k}-point vector of the second band index, $\textit{r}$, is given by $\textbf{k} - \textbf{Q}$.  Electron repulsion integrals at different $\textbf{Q}$ vectors are zero by symmetry, resulting in a reduction in the number of Cholesky vectors by a factor of $1/N_{k}$.
This in turn leads to a reduction in storage and computational costs by the same amount throughout the entire algorithm\citep{motta_kpoint}.
The AFQMC implementation in QMCPACK assumes a spin independent single-particle basis, which allows us to exploit time-reversal symmetry of the 2-electron integrals to further reduce storage.
For $\textbf{Q}$ vectors that satisfy $\textbf{Q} \ne -\textbf{Q}$ (this is not satisfied at the origin and at high symmetry points on the edge of the 1BZ), we have ${L^{\textbf{Q},\textbf{k}}_{sq,\gamma}}^{*} = {L^{-\textbf{Q},\textbf{k}-\textbf{Q}}_{qs,\gamma}}$, which requires us to store Cholesky vectors for either one of the $(\textbf{Q},-\textbf{Q})$ pair, but not both.

By using this $(\bQ,-\bQ)$ symmetry we can write the expressions for the force bias potential (see \cref{eq:fb})
\begin{align}
\bar{x}^{\pm}_{\textbf{Q}\gamma} & =  \sum_{\textbf{k}ar\sigma} \frac{c_{\pm}}{2} ( \mathcal{L}^{\sigma\textbf{Q},\textbf{k}}_{ar,\gamma} \ \mathcal{G}^{\sigma}_{(\textbf{k}a),(\textbf{k}-\textbf{Q}r)} \nonumber \\   
& \ \ \ \ \ \ \ \ \ \ \ \ \ \ \  \pm {\bar{\mathcal{L}}^{\sigma\textbf{Q},\textbf{k}}_{ra,\gamma}}  \mathcal{G}^{\sigma}_{(\textbf{k}-\textbf{Q}a),(\textbf{k}r)}  ), \label{eq:kp_vbias} 
\end{align}
and Hubbard--Stratonovich matrix as (see \cref{eq:vhs})
\begin{align}
V^{\mathrm{HS}}_{(\textbf{k}p),(\textbf{k}-\textbf{Q}r)} & = \frac{1}{2}  \sum_{\textbf{k}\gamma} [ L^{\textbf{Q},\textbf{k}}_{pr,\gamma} \left ( \tilde{x}^{+}_{\textbf{Q}\gamma} + \textit{i}\tilde{x}^{-}_{\textbf{Q}\gamma}  \right ) \nonumber \\
& \ \ \  + {L^{-\textbf{Q},\textbf{k}-\textbf{Q}}_{rp,\gamma}}^{*} \left ( \tilde{x}^{+}_{-\textbf{Q}\gamma} - i \tilde{x}^{-}_{-\textbf{Q}\gamma} \right ) ],\label{eq:kp_vhs} 
\end{align}
where in \cref{eq:kp_vhs} we introduced \mbox{${\tilde{x}}^{\pm}_{\textbf{Q}\gamma}=x^{\pm}_{\bQ\gamma}-\bar{x}^{\pm}_{\bQ\gamma}$} to simplify the notation. The expression for the force-bias potential, \cref{eq:kp_vbias}, uses the half-transformed Cholesky matrix, given by: $\mathcal{L}^{\sigma\textbf{Q},\textbf{k}}_{ar,\gamma} = \sum_{p} {[A^*_\sigma]_{(\textbf{k}p),(\textbf{k}a)}} L^{\textbf{Q},\textbf{k}}_{pr,\gamma} $, ${\bar{\mathcal{L}}^{\sigma\textbf{Q},\textbf{k}}_{ra,\gamma}} = \sum_{p} {[A^*_\sigma]_{(\textbf{k}p),(\textbf{k}a)}} {L^{\textbf{Q},\textbf{k}}_{rp,\gamma}}^{*} $, and the implementation assumes that the trial wave-function is block diagonal in $\textbf{k}$.  In \cref{eq:kp_vhs}, when $\textbf{Q} \ne -\textbf{Q}$, it is possible to perform a single contraction against the Cholesky matrix by using time-reversal symmetry and first summing over all $(+/-)$ contributions, $\tilde{x}_{\textbf{Q}\gamma} = \sum_{\pm} \left ( \tilde{x}^{\pm}_{\textbf{Q}\gamma} \pm \tilde{x}^{\pm}_{-\textbf{Q}\gamma} \right )$. 
Note that unlike the standard representation we do not form the intermediate structures $L_\pm$.

As discussed above, direct storage of the 2-electron integral tensor is typically prohibitive for systems with more than a few hundred orbitals. With this in mind, we implement the energy evaluation directly in terms of the Cholesky matrix, rather than with precomputed 2 electron integrals as is done in the Sparse representation. While this leads to a slightly higher computational cost in the energy evaluation, as we see below it allows us to reach systems with over 6000 basis functions. In terms of the Cholesky matrix, the expression for the local energy becomes:
\begin{align}
E_{L} & = \sum_{\textbf{k}aq\sigma} h_{(\textbf{k}a),(\textbf{k}q)} \mathcal{G}^{\sigma}_{(\textbf{k}a),(\textbf{k}q)} \nonumber \\
& + \frac{1}{2} \sum_{\substack{\gamma\textbf{Q}\textbf{k}\textbf{k}' \\ abrs\sigma\sigma'}} \mathcal{L}^{\sigma\textbf{Q},\textbf{k}}_{ar,\gamma} {\bar{\mathcal{L}}^{\sigma'\textbf{Q},\textbf{k}'}_{sb,\gamma}}  ( \mathcal{G}^{\sigma}_{(\textbf{k}a),(\textbf{k}-\textbf{Q}r)} \mathcal{G}^{\sigma'}_{(\textbf{k}'-\textbf{Q}b),(\textbf{k}'s)} \nonumber \\ 
& - \delta_{\sigma,\sigma'} \mathcal{G}^{\sigma}_{(\textbf{k}a),(\textbf{k}'s)} \mathcal{G}^{\sigma}_{(\textbf{k}'\textbf{Q}b),(\textbf{k}-\textbf{Q}r)}  ).\label{eq:kp_el}
\end{align}

Details about the efficient evaluation of these expressions are given below.

\subsection{Tensor Hyper-Contraction \label{sec:thc}}

Before finishing it is worth noting that neither the sparse representation nor the $k$-point representation is best suited for problems with larger mutli-atom unit cells or for systems with low symmetry (e.g. defective systems).
For cases such as these, we recently introduced the use of tensor hyper-contraction\citep{thc_1,thc_2,thc_3,lu_isdf,isdf_lin_1,isdf_lin_2,lee_thc_2020} (THC) based approaches in AFQMC\citep{malone_isdf}.
Briefly, in THC-AFQMC we write
\begin{equation}
    v_{pqrs} \approx \sum_{\mu\nu} \varphi^{*}_p(\br_\mu)\varphi_r(\br_\mu) M_{\mu\nu} (\varphi^{*}_s(\br_\nu)\varphi_q(\br_\nu))^{*}\label{eq:4ix_thc},
\end{equation}
where $\{\br_\mu\}_{\mu=1}^{N_\mu}$ is a set of real space `interpolating points' and
\begin{equation}
    M_{\mu\nu} = \int d\br d\br' \zeta_\mu(\br)K(\br,\br')\zeta^{*}_\nu(\br'),
\end{equation}
where $K(\br,\br')$ is the periodic Ewald potential.
In this form the computational cost of AFQMC can be formulated to scale cubicly with the system size with only a quadratic memory overhead.
Further implementation details can be found in Ref.\citenum{malone_isdf}. Note that THC can be combined with $k$-point symmetry to afford further savings in multi-atom cells.

\section{GPU Implementation\label{sec:gpu}}
In this section we describe in detail how we implemented AFQMC on GPUs.
As it is easy to write slow code on GPUs, we provide some insight on different optimization strategies we adopted.
We will pay particular attention to the $k$-point representation, which is the one whose performance shows the largest sensitivity to the details of the implementation.
In what follows we will distinguish between the standard implementation with dense and sparse linear algebra, as the \textit{dense} and \textit{sparse} representations respectively.   

We will focus solely on the simple test case of Carbon in the diamond structure (two atom unit cell) with a lattice constant of $3.6$\AA.
We used Goedecker-Teter-Hutter (GTH)\citep{GTH1996} (Pad{\'e}) type pseudo-potentials and the associated Gaussian basis sets,\citep{joost_gth_2007} as supplied by the CP2K software package\citep{gth_cp2k_2013}.
All calculations used restricted Hartree--Fock trial wavefunctions which were generated using the PySCF software package\citep{PYSCF}. The one- and two-electron integrals were also generated using PySCF using tools distributed freely with QMCPACK\citep{Kent_qmcpack_2020}. 

For the dense representation we constructed a supercell containing 2$\times N_{k}$ atoms at the Gamma point, leading to real orbitals which reduces the memory requirements and computational costs of the dense calculations by approximately a factor of 2.
For the sparse and $k$-point representation we used the $2$ atom cell and employed Brillouin-zone sampling using regular $\Gamma$-centered Monhorst-Pack grids\citep{monkhorst_pack}.
All AFQMC simulations were performed with a development version of QMCPACK\citep{qmcpack,Kent_qmcpack_2020,qmcpack_github}.
All AFQMC calculations were performed in single precision mode, where the Cholesky matrix is stored using single-precision floating point numbers and all associated tensor contractions are also performed in single precision, this results in very small modifications to the energies typically below 0.1 mHa/cell.
Cholesky factorizations were stopped when the magnitude of the largest error on the diagonal fell below $1\times 10^{-5}$ Ha. 

All of the data and scripts required to make the figures in this paper are available at Ref.~\citenum{materials_data} with additional details available in the Supporting Information\citep{supplement}. The supporting information also includes Refs\citenum{}.

\subsection{Scaling and Performance}

\begin{figure}
    \centering
    \includegraphics{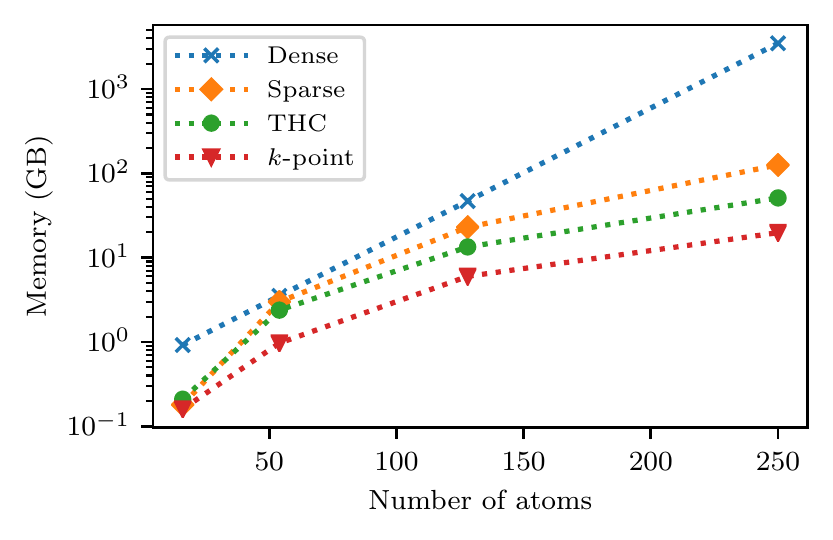}
    \caption{Memory (in GBs) needed to store static structures associated with the two-body Hamiltonian, as a function of the number of atoms in the calculation. This includes the Cholesky matrix, as well as pre-contracted two-electron integrals if used. The numbers in the figure correspond to Carbon with the GTH-DZVP basis set, which has 13 basis functions per carbon atom. \label{fig:memory}}
\end{figure}

\begin{figure}
    \centering
    \includegraphics{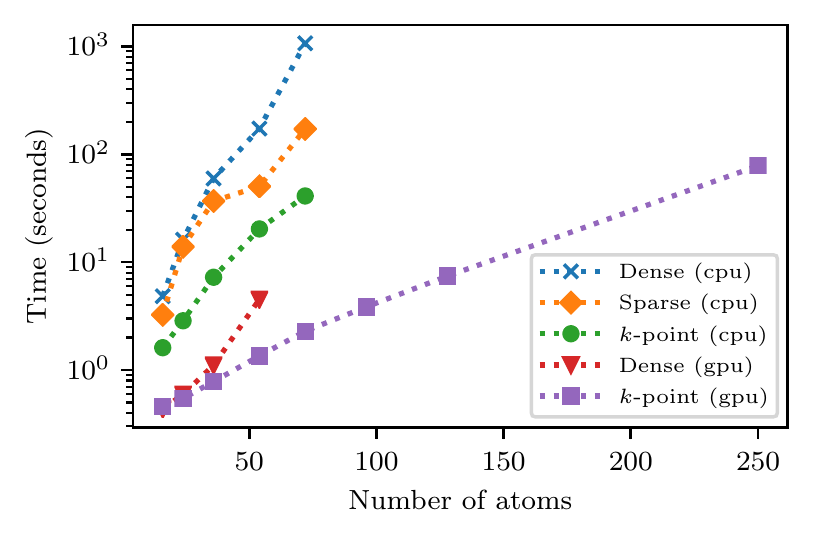}
    \caption{ Time (in seconds) per block of AFQMC calculation, as a function of the number of atoms in the calculation. A block in this case is defined as 20 iterations of the propagation step followed by an energy evaluation and walker orthogonalization. The `hybrid' method is used for propagation, which does not require the evaluation of the local energy. Both CPU and GPU times are reported. The numbers in the figure correspond to Carbon with the GTH-DZVP basis set, which has 13 basis functions per carbon atom. \label{fig:timing}}
\end{figure}

To begin we will briefly survey the memory consumption and performance of the various flavors of AFQMC described in the preceding sections.
\cref{fig:memory} shows the amount of memory needed to the store static data structures associated with the two-body Hamiltonian for the different representations.
The figure includes not only the Cholesky matrix and its half-rotated forms, but also pre-contracted two-electron integrals in the case of the sparse representation. As can be seen, the memory costs associated with the dense representation, which is mainly designed for isolated systems without symmetry, becomes quickly prohibitive as the system size grows since it makes no attempt to benefit from the sparsity of the Cholesky matrix resulting from translational symmetry. Both sparse and \textit{k}-point representations were built to directly benefit from this, resulting in much favorable scaling with system size. Notice that the sparse representation uses more memory because it stores pre-contracted two-electron integrals, which is found necessary in order to obtain reasonable performance in the energy evaluation.
The THC representation also offers favorable scaling, even though it doesn't take into account translational symmetry directly, and is built on a supercell representation of the simulation.
%This offers the possibility to study systems of comparable size even in the absence of translational symmetry, which would not be possible with either the sparse or \textit{k}-point representations. 

The exploitation of translational symmetry leads to great improvements in performance as well.
\cref{fig:timing} shows CPU and GPU execution times per block for AFQMC calculations in the GTH-DZVP basis. Here a block is defined as 20 iterations of the propagation step followed by an energy evaluation and walker orthogonalization.
The sparse representation has not been ported to GPUs yet in QMCPACK, so we only report CPU times. The THC representation only exits as a double precision CPU implementation for complex orbitals, so it was not included in the figure to avoid misleading conclusions.
In practice it should be somewhat slower than the sparse representation in CPUs \cite{malone_isdf}.

On the CPU we can see a significant improvement in performance for the \textit{k}-point representation for all system sizes, as can be expected, making the sparse representation practically obsolete for this type of calculations.
Notice that the sparse representation could be a leading alternative for the study of generic model Hamiltonians with highly simplified interactions, but it looses its utility in ab-initio periodic Hamiltonians with the introduction of the \textit{k-point} representation.
As the system size grows, the dense representation quickly becomes impractical and is only shown here for comparison purposes.

On the GPU, a slightly different picture emerges where the dense representation is competitive with the \textit{k}-point representation but only for small systems sizes, it quickly becomes significantly slower as the system size grows and eventually becomes impractical due to the much higher memory needs and the limited amount of memory in current devices (typically 16 GBs).
The \textit{k}-point representation, on the other hand, thrives in the GPU with a speedups on the order of x25-30 when comparing a node in Summit with 6 V100 GPUs and a node with a 36 core Intel Xeon processor. 

On a final note about memory usage, the AFQMC implementation in QMCPACK uses MPI-3 shared memory to keep a single copy of static data on each node, including all data structures associated with the 2-electron Hamiltonian. In addition, distributed memory implementations for all flavors of AFQMC are available which allow for the partitioning of the 2-electron Hamiltonian among a user defined number of nodes, including both multi-core and GPU architectures. As can be expected this leads to performance penalties due to additional communication, but enables calculations in cases where the memory requirements far exceeds the available memory on a node or GPU.  
To give a better idea of the applicability of AFQMC in systems with translational symmetry with this GPU implementation, we are able to study the current system in the GTH-TZVP basis with a $6\times6\times6$ $k$-point grid, which has 7344 basis functions and 1728 electrons. Using approximately 100 nodes in Summit for 24 hours, we are able to produce 1000 blocks of data which is more than enough to calculate accurate energies and properties. In this case, the Cholesky matrix was distributed among 3 Summit nodes (18 GPUs).

\subsection{Implementation on GPUs}

The performance of the \textit{k}-point representation in GPUs shown above opens the way for AFQMC studies of solids not previously possible. 
Unfortunately, extracting such performance from current GPU architectures is not a simple task and requires particular code design and careful optimization. For this reason, we discuss here details associated with the particular implementation of the method in QMCPACK.

GPUs have become ubiquitous in high performance computing and it looks like they will continue to be the path to higher performance in the next decade. While they offer significant increases in raw floating-point operations per second (FLOPS) compared to CPUs, this computing capacity can typically only be accessed by compute-intense operations, by highly parallelizable workflows, or by large numbers of independent tasks. Notice that data movement between GPU memory and CPU main memory is quite slow, so high performance almost always requires the entire calculation to be performed on the GPU, with little to no contributions from the CPU (only for task management operations). With these concepts in mind, we give the following guidelines for AFQMC implementations (some of these are already mentioned in Ref. \citenum{Shee_gpu_2018}): 1) all walkers should be operated on simultaneously when possible, 2) algorithmic steps should be implemented with matrix operations, always favoring level 3 BLAS operations over less compute-intensive versions and combining small multiplications in larger ones when possible, and 3) use of batched matrix operations when they can not be combined into a single larger operation. The latter recommendation is important in the case of the \textit{k}-point representation, where the typical calculation requires a large number (typically $N_{k}^{2}$ or $N_{k}^{3}$ ) of small operations (primitive cell dimensions) and there is no way to group them all into a single operation. Fortunately, batched BLAS and LAPACK implementations are provided by some GPU vendors like Nvidia and are also available through libraries like MAGMA \cite{tdb10}.

\begin{figure}
    \centering
    \includegraphics{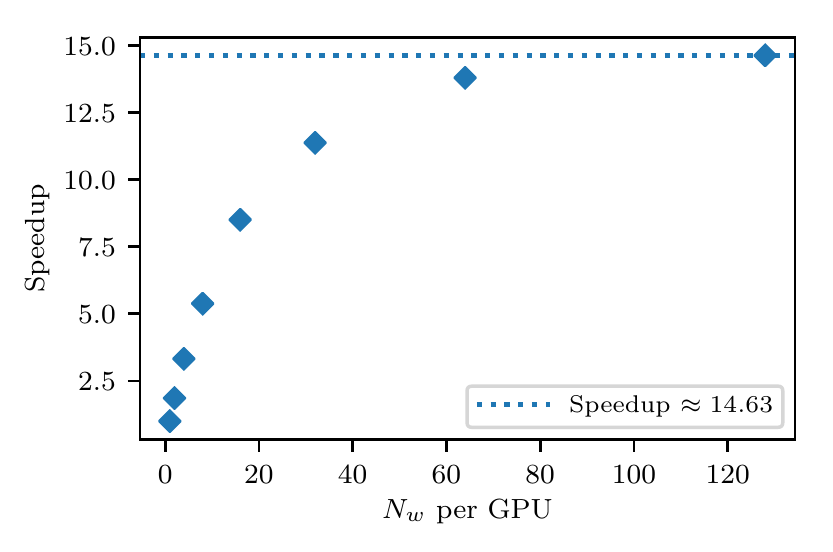}
    \caption{Measured speedup as a function of the number of walkers $N_w$ for the total simulation when batching over walkers. Here the speedup is defined as $(T(1)\times N_w) /T(N_w) $. Results correspond to simulations of Carbon in the GTH-DZVP basis set with a $3\times3\times3$ grid of \textit{k}-points, which corresponds to a simulation of 216 electrons in a single particle basis set with 702 functions.}
    \label{fig:batching_nw}
\end{figure}

\cref{fig:batching_nw} shows a representative speedup observed in the simulations of Carbon presented below as a function of the number of walkers in the simulation, $N_w$. We define the speedup as $(T(1)\times N_w) /T(N_w)$, the time to complete one block of simulation for single walker times $N_w$ divided by the time to complete a block with all walkers simultaneously. Specific results are shown for a simulation in the GTH-DZVP basis set with a $3\times3\times3$ \textit{k}-point grid, which corresponds to a simulation of 216 electrons in a single particle basis set with 702 functions. Code performance is poor for low walker counts since there is not enough work for full utilization of the GPU. As the number of walkers is increased, the GPU utilization increases until we reach a plateau on the speedup. Both the asymptotic speedup and the number of walkers needed to reach it will depend on system size, where smaller calculations will need more walkers to saturate the GPU. Larger walker populations lead to longer wall-clock times in general, so in practice a balance of the two can be made. Ultimately, we are limited by the amount of memory in the GPU since there is a memory cost proportional to the walker count on top of the memory required by static data structures. While an AFQMC simulation with 700 basis functions could be considered quite large by previous standards with CPU resources, it is routine now when executed in GPUs. 

\cref{fig:batching_nw} shows the speedup over a block of simulation which includes 20 propagation steps, walker orthogonalization and an energy evaluation. In order to obtain such speedup, multiple steps in the algorithm need to be carefully optimized.
The evaluation of the 2-body contribution of the local energy is typically the most time consuming step in the algorithm, fortunately the hybrid propagation scheme\cite{PhysRevB.80.214116} allows us to propagate walkers without an energy evaluation allowing us to reduce the frequency of evaluation to once per block. In the \textit{k}-point representation, the evaluation of \cref{eq:kp_el} is given by the following steps ($n$ is the walker index):
\begin{enumerate}
\item Reshape the half-rotated Green's function from  
\mbox{$\mathcal{G}^{\sigma}_{n,(\textbf{k}_1a),(\textbf{k}_2q)}  \rightarrow \mathcal{G}^{\sigma,\textbf{k}_1,\textbf{k}_2}_{naq}$} 
or, more explicity
\mbox{$\mathcal{G}[n,\bk_1,a,\bk_2,q]  \rightarrow \mathcal{G}[\bk_1,\bk_2,n,a,q]$} for each spin.
\item For each ($\sigma,\bQ,\bk,\bk'$):
\begin{enumerate}
\item Calculate $T^{1}_{nba\gamma} = \sum_{r} \mathcal{L}^{\sigma,\textbf{Q},\textbf{k}}_{a\gamma,r}  \mathcal{G}^{\sigma,\textbf{k}'-\textbf{Q},\textbf{k}-\textbf{Q}}_{nb,r}$,
\item Calculate $T^{2}_{nab\gamma} = \sum_{s} \mathcal{L}^{\sigma,-\textbf{Q},\textbf{k}'-\textbf{Q}}_{b\gamma,s}  \mathcal{G}^{\sigma,\textbf{k},\textbf{k}'}_{na,s}$,
\item Accumulate the exchange contribution: \mbox{$E_{L}(n) \mathrel{+}= -\frac{1}{2} \sum_{ab\gamma} T^{1}_{nba\gamma} T^{2}_{nab\gamma}$},
\item Accumulate $v_{jn\gamma} \mathrel{+}= \sum_{a} T^{j}_{naa\gamma}$
\end{enumerate}
\item Calculate the Coulomb contribution: \mbox{$E_{L}(n) \mathrel{+}= \frac{1}{2} \sum_{\gamma} v_{1n\gamma} v_{2n\gamma}$}
\end{enumerate}

\begin{figure}
    \centering
    \includegraphics{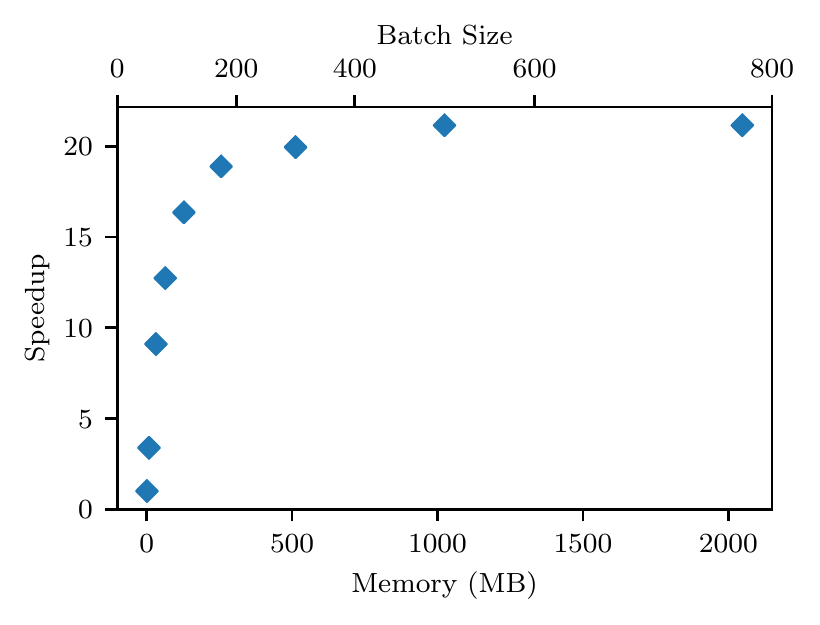}
    \caption{Measured speedup as a function of the batch size (top axis) and memory usage (bottom axis) for the energy evaluation. Here the speedup is defined as the time taken to perform an energy evaluation for a given batch size measured with respect to sequential execution (batch size of 1). Results correspond to simulations with 24 walkers in the GTH-DZVP basis set with a $3\times3\times3$ grid of \textit{k}-points.}
    \label{fig:batching_mem}
\end{figure}

Notice that in steps 2(a) and 2(b) we use a transposed form of the half-transformed Cholesky matrix, which allows us to perform these steps with a dense matrix-matrix multiplication (GEMM). Each of the operations in step 2 are very small compared to the overall cost of the energy evaluation, performing these operations sequentially would lead to very poor performance even when all walkers are processed simultaneously. We use batched operations, which allow us to process simultaneously an arbitrary number of terms in the sum over ($\sigma,\bQ,\bk,\bk'$).
Steps 1, 2(c), 2(d) and 3 are performed with a custom-made CUDA kernels which are written to process multiple batches and walkers concurrently. The number of terms of the triple sum over \textit{k}-points which are processed simultaneously is determined by the size of the memory buffer available for the computation (needed to store $T^1$, $T^2$), which can be controlled by the user. The larger the buffer space, the higher the speed of evaluation. \cref{fig:batching_mem} shows the speedup obtain in simulations with 24 walkers using the GTH-DZVP basis set and 27 \textit{k}-points. The speedup is defined with respect to sequential execution of step 2, but processing all walkers simultaneously. A can be seen, speedups as large as x20 are observed in this case. A similar implementation is used for the force-bias (Eq. \ref{eq:kp_vbias}) and the Hubbard-Stratonovich (Eq. \ref{eq:kp_vhs}) potentials, where transposed Green functions are used and nested sums over \textit{k}-points are performed concurrently using batched operations. 

\begin{figure}
    \centering
    \includegraphics{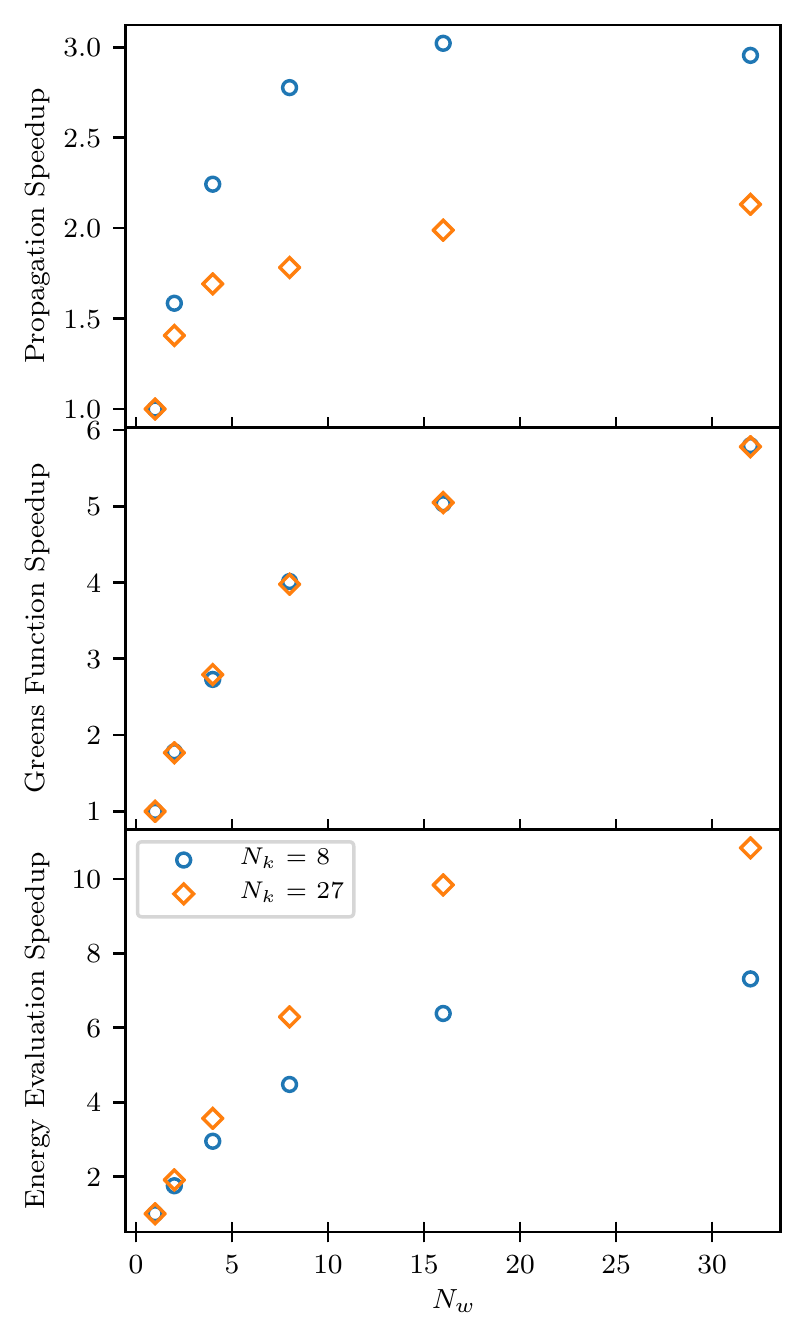}
    \caption{Measured speedup for walker propagation, Green's function evaluation and energy evaluation as a function of the number of walkers in the batch, $N_w$. Results correspond to simulations in the GTH-DZVP basis set for 2 different number of \textit{k}-points.}
    \label{fig:batching_timing}
\end{figure}

Other operations can significantly benefit from batched processing including walker propagation, orthogonalization and Green function evaluations. \cref{fig:batching_timing} shows the speedup obtained by processing multiple walkers simultaneously in: 1) the application of the Hubbard-Stratonovich potential, 2) the evaluation of the Green's function (\cref{eq:gfhrot}) and 3) in the energy evaluation (\cref{eq:kp_el}). The first two are common to all representations, the latter being specific to the \textit{k}-point representation. Two system sizes are shown, corresponding to simulations with 8 and 27 \textit{k}-points. The speedup of the various operations depends differently on system size, but all three operations benefit significantly from larger walker batches. In QMCPACK, the default behavior is concurrent processing of all walkers in the population. Since concurrent processing of walkers invariably leads to larger memory requirements, the user can control batch sizes in our implementation if desired, which can be useful in simulations with system sizes where memory on the GPU becomes limited. 

\section{Results and discussion \label{sec:res}}
\begin{figure}
    \centering
    \includegraphics{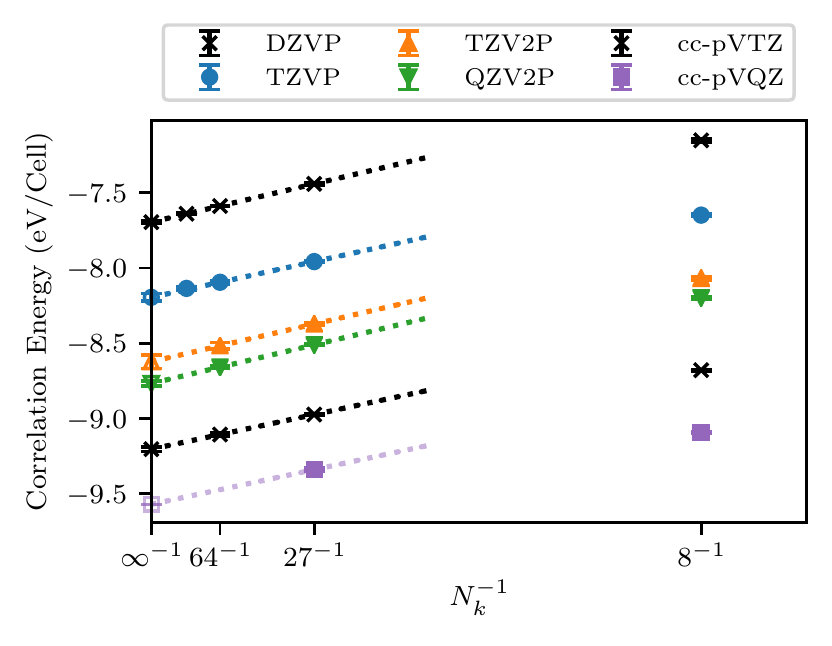}
    \caption{Convergence of AFQMC correlation energy with the number of $k$-points ($N_k$) for different basis set sizes. The dashed lines represent a linear two-point extrapolation of the \mbox{$N_k=(27,64)$} data points. The extrapolated values are plotted at $N_k=\infty$ with error bars accounting for the effect of the AFQMC statistical error bars on the extrapolation. For the cc-pVQZ we plot a tentative extrapolated value (light colored purple square) using a the size correction from the cc-pVTZ data.}
    \label{fig:ecor_conv}
\end{figure}

%Finally, we use our new capabilities to study the cohesive energy of Carbon at ambient conditions, paying particular attention to its convergence with respect to system size and basis set. We obtain a cohesive energy in excellent agreement with experimental measurements when both dependencies are carefully taken into account.    
%In this section we study Carbon in the diamond structure, a prototypical covalently-bonded solid. Solid carbon has been studied in the past in numerous occasions with various quantum many-body methods including: MP2 \cite{something}, RPA \cite{something}, CCSD \cite{something}, CCSD(T) \cite{something}, AFQMC \cite{something}, DMC \cite{something}, among others. 
%We present scaling and performance comparisons of the various representations of the Hamiltonian in AFQMC described above, as well as show the current reach and capabilities of the implementation of the method in QMCPACK, particularly in GPU architectures where we obtain dramatic improvements in performance compared to traditional CPU-only architectures. 
In this section we present our results for the cohesive energy of Carbon in the diamond structure.
%We use the PySCF \cite{something} software package to generate all Hartree-Fock, MP2, CCSD and CCSD(T) calculations presented in the article.
%We use the GTH \cite{something} pseudo-potential and basis sets for Carbon. Notice that the GTH basis sets distributed with their pseudo-potentials are not of the correlation consistent form, a fact that will be important in our discussions below.
In \cref{fig:ecor_conv} we plot the convergence of the AFQMC correlation energy with the number of $k$-points sampling the Brillouin zone $N_k$.
As can be seen the data follows a smooth $1/N_k$ behavior (equivalent to $1/N$).
Moreover, we can see that extrapolating to the thermodynamic limit $(N_k\rightarrow\infty)$ can be safely perfomed using just the $3\times3\times3$ and $4\times4\times4$ $k$-point grids. 
Interestingly for this system, the size dependence of the correlation energy is quite uniform across basis sets, with the AFQMC data for different basis sets being roughly parallel to one another\citep{supplement}. 
This suggests that a size correction can be approximately obtained from a smaller basis set and applying this to CBS AFQMC correlation energies obtained in a more affordable $k$-point mesh. 
What is also obvious from \cref{fig:ecor_conv} is the fact that the GTH basis sets were not developed for correlated calculations\citep{delben_2012_pmp2}.
Although at first glance the AFQMC correlation energy appears to be converging when moving from the TZV2P to the QZV2P basis set, in reality it is far from the case. 

To demonstrate this we generated a modified correlation-consistent basis set for the GTH pseudo-potential of Carbon in the spirit of Dunning basis sets, which are widely used in quantum chemistry\citep{Dunning1989}. 
To do so we took the valence states of the GTH-TZVP basis and added virtual states from the cc-pVXZ basis of Carbon from the BFD pseudo-potential\cite{bfd_pseudos_07}, where $X$ is the cardinality of the basis set ($X=$D,T,Q).
This produced a modified correlation consistent basis set that proved to be good enough to generate satisfactory basis set extrapolations of the correlation energy in the solid and in the isolated atom\cite{supplement}.
We see from \cref{fig:ecor_conv} that huge correlation energy gains are possible if appropriate basis sets are used, although the cost of using such basis sets can be prohibitive.

\begin{figure}
    \centering
    \includegraphics{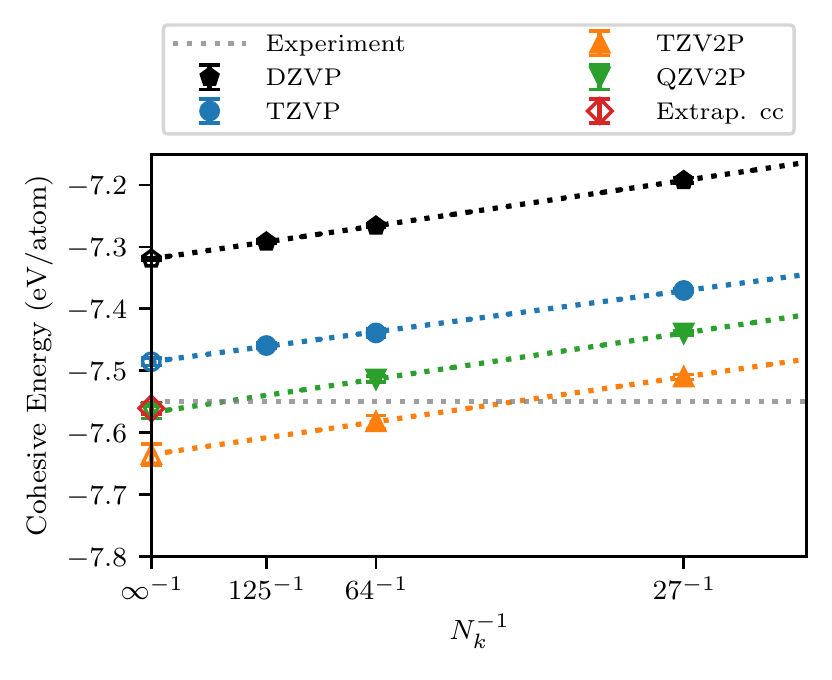}
    \caption{Convergence of AFQMC cohesive energy with the number of $k$-points ($N_k$) for the GTH basis sets. The dashed lines represent a linear extrapolation of the \mbox{$N_k=(27,64,125)$} data points for the DZVP and TZVP basis sets while we used \mbox{$N_k=(27,64)$} for TZV2P and QZV2P. Cohesive energies for the GTH basis sets are counterpoise corrected. The extrapolated values are plotted at $N_k=\infty$. Also plotted is the correlation consistent CBS AFQMC cohesive energy which has been size corrected as described in the main text (red diamond, labelled `Extrap cc.'). The experimental value has been corrected for zero point effects\citep{schimka2011improved}.}
    \label{fig:cohesive}
\end{figure}

Despite these difficulties we find that energy differences computed using these basis sets often yield quite satisfactory results compared to experiment.
In particular we plot in \cref{fig:cohesive} the cohesive energy $\Delta E = E_{\mathrm{solid}}-E_{\mathrm{atom}}$ of diamond using the GTH basis sets.
For these calculations the energy of the solid and atom were calculated using the same basis set and pseudopotential.
For the atomic calculation we used a counterpoise correction\citep{boys_1970_counterpoise,maschio_2010_pmp2,delben_2012_pmp2} (CP)\citep{supplement}.
For the solid we separately converged the Hartree--Fock energy to the $N_k\rightarrow\infty$ limit using the spherically truncated Coulomb potential\citep{spencer_sphcoul_08,supplement}.
We see that the cohesive energy converges non-monotonically with basis set size, although the scatter in the data is on the order of 0.1 eV / atom which could easily be accounted for by different systematic effects, such as a poor treatment of the atom or solid in the larger basis sets.

We also plot the cohesive energy computed using our modified correlation consistent basis sets, where we separately extrapolated the solid to the CBS and system size limit, and extrapolated the atom to the CBS limit. 
This procedure is labelled as Extrap cc. in \cref{fig:cohesive}.
In this case we extrapolated the AFQMC correlation energy for the $3\times3\times3$ $k$-point mesh to the CBS limit using a two point extrapolation ($X=T,Q$).
We then applied a size correction to the correlation energy calculated from the cc-pVTZ basis set\citep{supplement}.
Again we separately converged the Hartree--Fock energy to the CBS and infinite $k$-point limit.
For the atom we extrapolated the correlation energy in the same correlation-consistent sets and pseudopotentials to the CBS limit\citep{supplement}.
As we can see this procedure agrees exceptionally well with experiment, although this level of agreement is probably fortuitous.
Convincingly, both approaches produce results that are within roughly 0.02 eV of each other.
The results are summarized in \cref{tab:cohesive_energy} where we also provide cohesive energies computed without the counterpoise correction and compare to previously published CCSD and CCSD(T) results.
We see that CCSD differs from the experimental result by between 0.5 and 0.25 eV / atom, whilst our AFQMC results agree well with CCSD(T) as well as experiment.
%\todo{We attribute this to correlation energy errors in the solid phase (see Supporting Information), with the CCSD error relative to CCSD(T) being on the order of x mHa / Cell.
%In contrast, the CCSD energy for the atom is far more accurate leading to an underestimation in the cohesive energy. I'm completely bullshitting here, I don't know if this is actually the case}. 
% MAM: we should not make comments about CCSD accuracy, unless we are sure.
This result adds to the growing body of literature placing AFQMC near par with CCSD(T) in terms of accuracy for relative energies\citep{lee_st_afqmc_2020}, albeit at a much reduced cost in terms of wall time.

\begin{table}
\begin{tabular}{llr}
\hline
Method &     Basis Set & Cohesive Energy \\
\hline
CCSD \citep{mcclain2017gaussian}& TZVP(CP) & -7.01 \\
CCSD \citep{booth2013towards} & PW  &      -7.295 \\
CCSD(T) \citep{booth2013towards} & PW  & -7.545 \\
\hline
AFQMC  &     DZVP &       -7.362(3) \\
 &   DZVP(CP) &       -7.319(3) \\
 %&   \hline
&     TZVP &       -7.549(6) \\
 &   TZVP(CP) &       -7.485(5) \\
 &    TZV2P &        -7.69(2) \\
 &  TZV2P(CP) &        -7.64(2) \\
&     QZV2P &       -7.587(8) \\
&   QZV2P(CP) &   -7.567(8)\\
 %   \hline
 & Extrap. cc &        -7.56(1) \\
 \hline
 %\hline
 Experiment \citep{schimka2011improved} &        & -7.545 \\
\hline
\end{tabular}
\caption{Cohesive energies of Carbon computed using CCSD, CCSD(T) and  AFQMC compared to experiment. Energies are in eV/atom and experimental results have been corrected for zero-point effects. Note that we used the same GTH basis sets and pseudopotential as those used for the coupled cluster results of Ref.\citenum{mcclain2017gaussian}. The coupled cluster results of Ref.\citenum{booth2013towards} used the PAW framework.  \label{tab:cohesive_energy}}
\end{table}

%\begin{table}
  %\caption{Cohesive energies of a series of systems calculated using AFQMC. $a$ is the lattice constant used in our calculations. $E_\text{coh}^\text{expt.}$ are the experimental values listed in Table V of Ref.~\cite{HSEsol2011} including the zero-point-energy correction. AFQMC values needed }
  %\label{tbl:Ecoh}
  %\begin{tabular}{llll}
    %\hline
    %& $a$ ($\AA$)  & $E_\text{coh}^\text{AFQMC}$ (eV/atom) & $E_\text{coh}^\text{expt.}$ (eV/atom) \\
    %\hline
    %Li (bcc) & 3.51   & 1.81--2.07 & 1.66 \\
    %Al (fcc) & 4.0495 &            & 3.43 \\
    %C (diamond) & 3.56683 & 7.47--7.76 & 7.54 \\
    %LiH (B1)    & 3.979 & 5.03     & 4.97 \\
    %LiF (B1)    & 3.972 & 9.59     & 8.91 \\
    %\hline
  %\end{tabular}
%\end{table}

\section{Conclusions \label{sec:conclusions}}
In this paper we outlined how recent algorithmic developments along with new computing architectures can dramatically extend the scope of AFQMC simulations of solids. 
By reformulating the algorithm to explicitly account for $k$-point symmetry we showed it possible to efficiently use modern GPUs by making extensive use of batched linear algebra.
With these developments we were able to simulate system sizes outside the reach of the standard formulation of AFQMC, without using massive computational resources.
For example, the $4\times4\times4$ simulations for the TZVP basis set, corresponding to 512 electrons in 2176 basis functions, could be run on 8 nodes on Summit for 4 hours to get results with errorbars on the level of 0.2 mHa / Cell.
Using these developments we found it possible to systematically reach the thermodynamic and CBS limit for relatively simple systems such as Carbon.
We found that cell sizes ($k$-point meshes) of $3\times3\times3$ and $4\times4\times4$ were sufficient in order to extrapolate to the thermodynamic limit, an observation that might be transferable to many similar two- or four-atom cells.
These system sizes should be routine to perform on relatively modest computational resources.
We also demonstrated the need to use appropriately constructed correlation consistent basis sets if accurate correlation energies are desired.
With these developments we hope that AFQMC can be more broadly and routinely applied to solid state systems, with this work serving as a stepping stone in this direction.
\begin{acknowledgments}
We would like to thank Mario Motta for helpful discussions and Qiming Sun for assistance with running PySCF. 
We thank Joonho Lee for his insistence on our reporting of counterpoise corrected finite basis set cohesive energies and for other helpful criticism.
This work was performed under the auspices of the U.S. Department of Energy (DOE) by LLNL under Contract No. DE-AC52-07NA27344. Funding support was from the U.S. DOE, Office of Science, Basic Energy Sciences, Materials Sciences and Engineering Division, as part of the Computational Materials Sciences Program and Center for Predictive Simulation of Functional Materials (CPSFM). Computer time was provided by the Livermore Computing Facilities and Oak Ridge Leadership Computing Facility. An award of computer time was provided by the Innovative and Novel Computational Impact on Theory and Experiment (INCITE) program. This research used resources of the Oak Ridge Leadership Computing Facility, which is a DOE Office of Science User Facility supported under Contract DE-AC05-00OR22725. FDM and MAM acknowledge support for the GPU implementation of AFQMC in QMCPACK from the Exascale Computing Project (17-SC-20-SC), a collaborative effort of the U.S. Department of Energy Office of Science and the National Nuclear Security Administration.
\end{acknowledgments}

\bibliography{refs}

\begin{thebibliography}{85}
\bibitem [{pw_()}]{pw_footnote}%
  \BibitemOpen
  \href@noop {} {}\bibinfo {howpublished} {Note that plane wave approaches for
  AFQMC\citep{Zhang_phaseless,suewattana_pw_afqmc,ma_multiple_proj} can avoid
  the memory cost associated with storing the two-electron integrals, necessary
  when employing gaussian orbitals, at the cost of potentially requiring a much
  larger plane wave basis set to converge quantities.}\BibitemShut {Stop}%
\bibitem [{mat()}]{materials_data}%
  \BibitemOpen
  \href@noop {} {}\bibinfo {howpublished} {All simulation data is available at
  [url to be inserted].}\BibitemShut {Stop}%
\end{thebibliography}

\end{document}